# Synthetic ablations in the *C. elegans* nervous system

Emma K. Towlson[1,2] and Albert-László Barabási[1,3,4,5]

*1. Center for Complex Network Research and Department of Physics, Northeastern University, Boston, Massachusetts 02115, USA*

*2. Media Laboratory, Massachusetts Institute of Technology, Cambridge, Massachusetts 02139, USA*

*3. Center for Cancer Systems Biology, Dana Farber Cancer Institute, Boston, Massachusetts 02115, USA*

*4. Department of Medicine, Brigham and Women's Hospital, Harvard Medical School, Boston, Massachusetts 02115, USA*

*5. Center for Network Science, Central European University, H-1051 Budapest, Hungary*

**Abstract**

*Synthetic lethality*, the finding that the simultaneous knockout of two or more individually non-essential genes leads to cell or organism death, has offered a systematic framework to explore cellular function, and also offered therapeutic applications (1,2). Yet, the concept lacks its parallel in neuroscience – a systematic knowledge base on the role of double or higher order ablations in the functioning of a neural system. Here, we use the framework of network control (3–10) to systematically predict the ablation of neuron pairs and triplets. We find that surprisingly small sets of 58 pairs and 46 triplets can reduce muscle controllability, and that these sets are localised in the nervous system in distinct groups. Further, they lead to highly specific experimentally testable predictions about mechanisms of loss of control, and which muscle cells are expected to experience this loss.

**Introduction**

``Synthetic lethality'' (11), a term well established in cell biology, refers to the phenomenon whereby the deletion of individual genes is tolerated by an organism, but the deletion of the combination is lethal. Such synthetic lethal pairs are of particular interest due to their importance for cancer therapies (1,11). Double or higher order gene deletions can also affect the growth rate or other quantitative traits of a cell (12), and they are of particular interest if the outcome is not explained by the simple summation of the phenotypes of the individual knockouts (13). Studies in yeast (*Saccharomyces cerevisiae*) (14) have found that digenic interactions tend to be enriched with single genes known to affect fitness, i.e. most are positive or negative interactions of essential genes (15,16). Trigenic interactions have been shown to be weaker in magnitude than digenic interactions, and spread across many different genes/distant bioprocesses, although functionally related genes are hubs on the trigenic network (17,18).

Taken together, double and higher knockouts have offered a systematic tool to explore cellular systems, and have led to multiple mechanistic insights, as well as therapeutic applications. Despite its widespread use and value in cell biology, the systematic study of higher order neural ablations are lacking in neuroscience. Certainly, many studies in *C. elegans* have investigated the effects of single neuronal *class* ablations (two to 13 individual neurons), and much fewer the effects of two or more class ablations (19). These simultaneous ablations tend to be small sets of neurons targeted by a precise hypothesis or query, and in the context of specific functions of interest. Early work includes probing the circuitry behind touch sensitivity (20,21), finding, for example, that the simultaneous ablation of AVA and AVD leads to a loss of the worm's ability to move backwards, an effect not observed with the ablation of either class alone. Such experiments were collectively able to identify two pathways for anterior touch induced locomotion, and one for posterior touch. Multiple ablations within the pharynx revealed the surprising result that no neuron in the pharyngeal nervous system is necessary for pharyngeal pumping, and only one (M4) is required for the animals to grow into fertile adults (22). Further, the pharyngeal neurons may be divided into three functional groups based on the behavioural effects of ablating them. A study which systematically ablated combinations of chemosensory neurons demonstrated that the simultaneous ablation of the four classes ADF, ASG, ASI, ASJ produces larval worms which enter the dauer stage – a specialised form resistant to harsh conditions – regardless of environment (23). Smaller subsets of these neuronal ablations did not lead to the same effect, revealing the functional interdependence of these classes. Ablation of the AVL and DVB neurons, individually and as a pair, is moderately to severely detrimental to the defection cycle, with the most pronounced effect produced by the double ablation (24). This highlights an amount of redundancy in their function, and given the dependence of enteric muscle contractions on these GABA-ergic neurons, provides evidence that GABA may be a *stimulatory* as well as inhibitory neurotransmitter.

Much of the reason for such carefully targeted and class-based queries – instead of systematically exploring the whole space of individual neurons – is the tractability of performing so many experiments. Yet, it is likely that systematic data on the outcome of double and higher ablations could offer a resource as useful to the community as higher order knockouts are to cell biology. In this paper we set out to do just that – use the tools of network control to offer a complete set of predictions on double and triple ablations' effect on locomotion. Indeed, a comprehensive set of theoretical predictions could also guide future experiments, and inform of mechanistic network effects in neural circuits. The predictions can expose network level redundancy and robustness, and potentially even guide restoration of lost function. The emergence of genetic ablation methodologies and precise optical targeting (25) promise tools which may make systematic double and triple ablations possible in the near future.

Synthetic lethality is necessarily defined in relation to some phenotype: whether a genetic component is essential to the function in question. In single-cell organisms, the metric of fitness is usually growth or lethality. In neural systems, one could consider multiple such phenotypes. Here, we illustrate the value of systematic ablation studies in neuroscience by exploring the space of *neuronal* functional interactions in the *C. elegans* nervous system and their role in locomotion. This is made possible by the development of network control (3–5), a theoretical framework that helped frame the locomotor response of *C. elegans* to sensory input as a target control problem (26). This helped to identify 20 neurons which when ablated individually are predicted to lead to a loss of muscle control – see Supplementary Table S3. We hypothesise that due to network effects, there would be non-essential neurons as categorised by this approach which, when simultaneously ablated, would have an impact on controllability of the muscles. This *pair* (or *triplet*) would then be essential for locomotion. Therefore, using the same theoretical framework, here we systematically ablated *in silico* each possible neuronal pair, and each possible neuronal triplet, and examined the effects, if any, on muscle controllability. Our results point to highly localised, and highly specific, double and triple neuronal interactions in locomotor control. We identify the small groups of muscles affected by these interactions, together with the few network-level mechanisms behind the loss of control.

**Results**

For each possible pair and triplet of neurons, we performed a structural controllability analysis with these neurons and their connections removed from the wiring diagram. We employed a target control approach, with the mechanosensory neurons for gentle touch as input nodes ({ALML, ALMR, AVM} for anterior touch, or {PLML, PLMR} for posterior touch) and the 95 body muscle cells as output nodes (see Figure 1 and Materials and Methods). This allowed us to quantify the number of muscles which can be independently controlled via the application of suitable input signals to the input neurons. Any deviation from the healthy case signals an impact on controllability caused by the ablation. Given that there are 279 non-pharyngeal neurons in the *C. elegans* nervous system (not excluding any sensory input for the purposes of this illustrative calculation), we have a set of 279 neurons from which to draw neurons for candidate ablations. We therefore tested $^{279}C_2 = 38,781$ double neuronal ablations, and $^{279}C_3 = 3,580,779$ triple neuronal ablations.

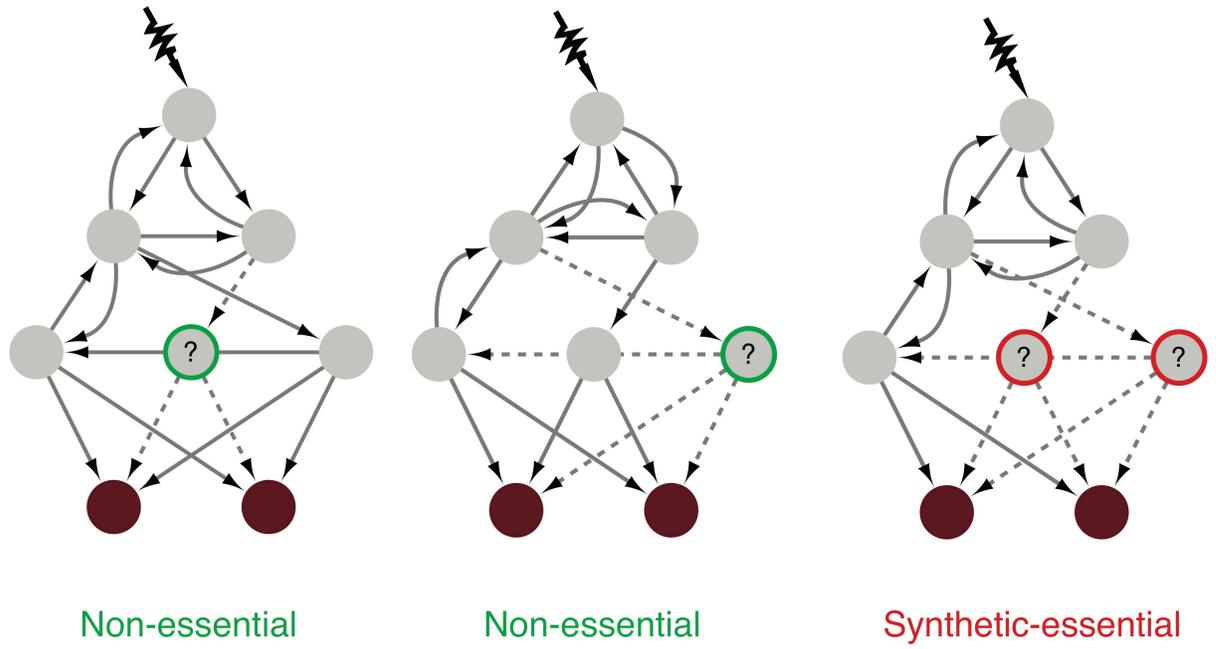

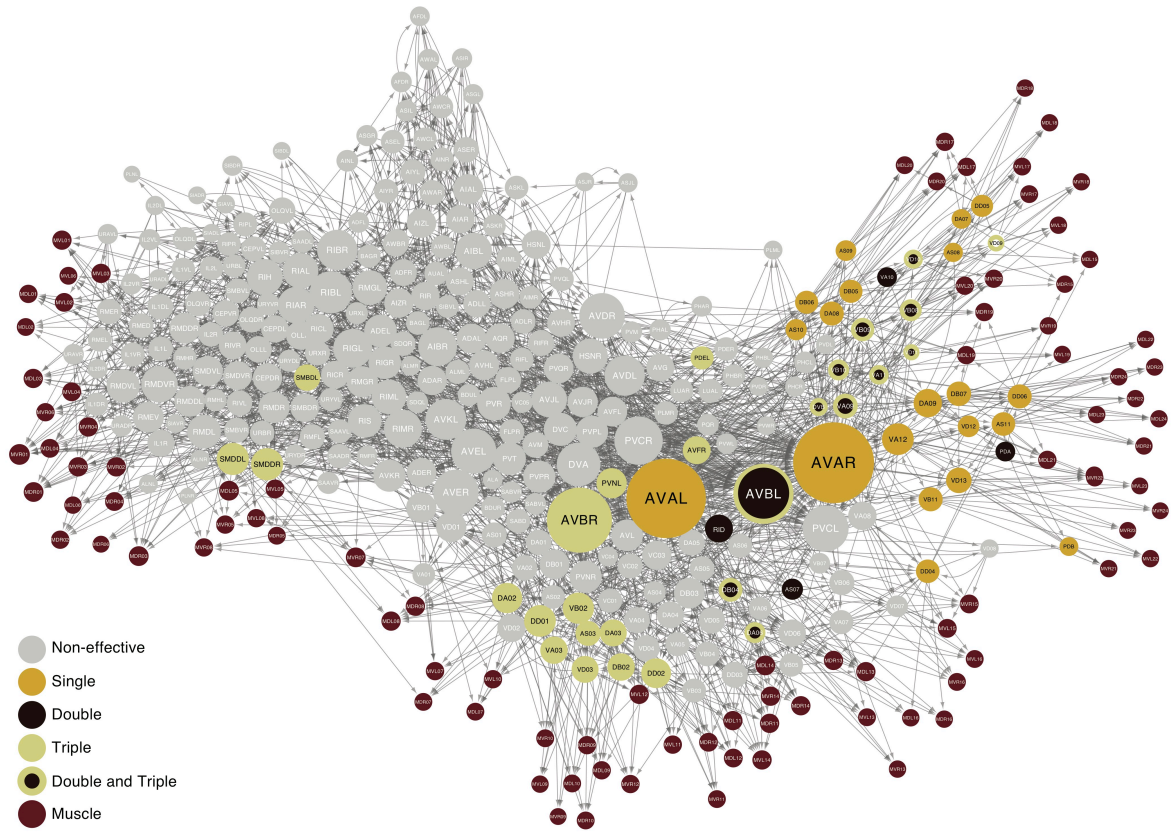

**Figure 1: Synthetic essentiality in double and triple neuronal ablations.** (a) Individual neurons are systemically ablated from the wiring diagram, and the impact on muscle controllability assessed. Neurons which do not lead to a loss of control are deemed 'non-essential'. In a double ablation, two neurons may be non-essential individually, but when they are both ablated, we predict a reduction in the number of controllable muscles and the *pair* is essential. (b) Predicted essential single, pairs of, and triplets of neurons in the *C. elegans* connectome. Mustard nodes denote individual neurons predicted to be essential as per (3). Neurons involved in a pair predicted to be synthetic essential are coloured dark brown, and those in a triplet are coloured yellow-green; if a neuron is part of a pair and a triple, it is coloured dark brown and yellow-green. Muscles are in red-brown and non-essential neurons in grey, and all cells are sized by nodal degree.

*Synthetic pairs.*

We find that 5428 of the 38,781 possible double ablations lead to a loss of control, while the remaining 33,353 have no effect on locomotion; see Table 1 and Supplementary Tables S1 and S4. Of the 5428 pairs that do affect controllability, 190 comprise neuron pairs whose members were already predicted (and validated) to be essential via single ablations ($E_{single}E_{single}$), 5180 contain one previous neuron and one individually non-essential ($E_{single}N_{single}$), and one is the trivial case of the removal of the pair of input neurons in the case of posterior touch (meaning no external control signal is received by the network). The remaining 57 pairs have no overlap with the set of predictions from single ablations ($N_{single}N_{single}$), and thus constitute synthetic essential pairs. They involve 16 distinct neurons, most of which are ventral motorneurons, and are predicted to affect a small set of muscles in the ventral posterior section of the body (see SI Section III).

We also note a single case of enhancement of control loss, i.e. a negative interaction (see SI Section I). In genetic knockouts, a *negative* interaction enhances the detrimental effect of a knockout, while a *positive* interaction diminishes it. AS11 is predicted to be essential individually, leading to a reduction in control from 89 to 88 muscles. (Note that the AS class as a whole has been experimentally implicated in locomotory coordination (27).) When RID is simultaneously ablated, we predict that this number further reduces to 87 independently controllable muscles.

Interestingly, for each ablated neuron, the number of independently controllable muscles reduces by a maximum of one. Given that 89 muscles are found to be controllable in the healthy worm (3), a double ablation leads to at most a reduction of two, to 87 controllable muscles.

*Mechanisms.*

Neuronal ablations affect controllability by changing the number of linearly independent control signals which arrive at the muscles from the input neurons. In our original study, we demonstrated that single ablations can achieve the same effect by reducing the numbers of the sets of motor neurons directly connecting to sets of muscles (3). Here, we find that double ablations act one layer higher – they reduce the number of linearly independent control signals received by the *motor neurons*, causing a knock-on effect of further constricting the possible number that can ultimately arrive at the muscles.

To be specific, we identified three distinct mechanisms which lead to a loss of control in double ablations (see Figure 2(a) and Figure 3). Let us label the muscles as Layer 0 cells, and we label any neuron with a direct connection to a muscle a Layer 1 neuron. Finally, neurons that are a path length of two away from a muscle form Layer 2, and so on, such that the Layer number of a neuron is the length of the shortest path from that neuron to a muscle cell. Figure 2(a) illustrates the nature of the three mechanisms that lead to loss of control in synthetic essential neuron pairs.

Mechanism 1 (40 synthetic essential pairs): Figure 2(a) first shows a network with two Layer 0 nodes, four Layer 1 nodes, and two Layer 2 nodes. In this network, however, the Layer 2 nodes do not fully connect to the Layer 1 nodes, rather splitting them into two groups of two, each of which receives one control signal. The removal of two nodes from one of these groups disconnects one control signal, and control is lost over one of the Layer 0 nodes.

Figure 2(b) illustrates the ablation of the pair {VA09, VA10}, which we find to display synthetic essentiality via this mechanism. The two muscle cells MVL18 and MVR18 connect to only four neurons, VA09, VA10, VB08, and VD09. These four neurons are likely to receive only three independent control signals, meaning correlation is expected in their activity levels. If either VA09 or VA10 is ablated, at least two independent control signals remain, which is enough to control the two muscles. But if both VA09 and VA10 are ablated, it is likely that only one independent control signal remains, and control is lost over one of the muscle cells MVL18 or MVR18.

Mechanism 2 (1 synthetic essential pair): The second network in Figure 2(a) has two Layer 0 cells, two Layer 1, and three Layer 2 nodes. If we remove a Layer 2 neuron, there are still enough independent control signals received by the Layer 1 nodes to control the muscles in Layer 0. But if we remove a second Layer 2 neuron, only one control signal is received by the two Layer 1 neurons, and consequently only one Layer 0 neuron can be controlled. The pair which is governed by this mechanism is {AVBL, DVB}.

Mechanism 3 (16 synthetic essential pairs): The third network in Figure 2(a) has two muscle cells, which connect to three Layer 1 neurons, which receive input from two Layer 2 neurons. Since there are more Layer 1 neurons – each of which receives a linearly independent signal – than muscles, each of the muscles is independently controllable (3). If one Layer 1 neuron is removed, the muscles are still controllable, as there is still one independent signal from each Layer 1 neuron for each one muscle. But if a Layer 2 neuron is *also* removed, only one independent control signal arrives at the two Layer 1 neurons. This means there is only one independent control signal to control two muscles, and control is therefore lost over one of them. Synthetic essential pairs governed by Mechanism 3 comprise either AVBL or DVB and a ventral motor neuron, such as {AVBL, VB10} and {DVB, VD10}.

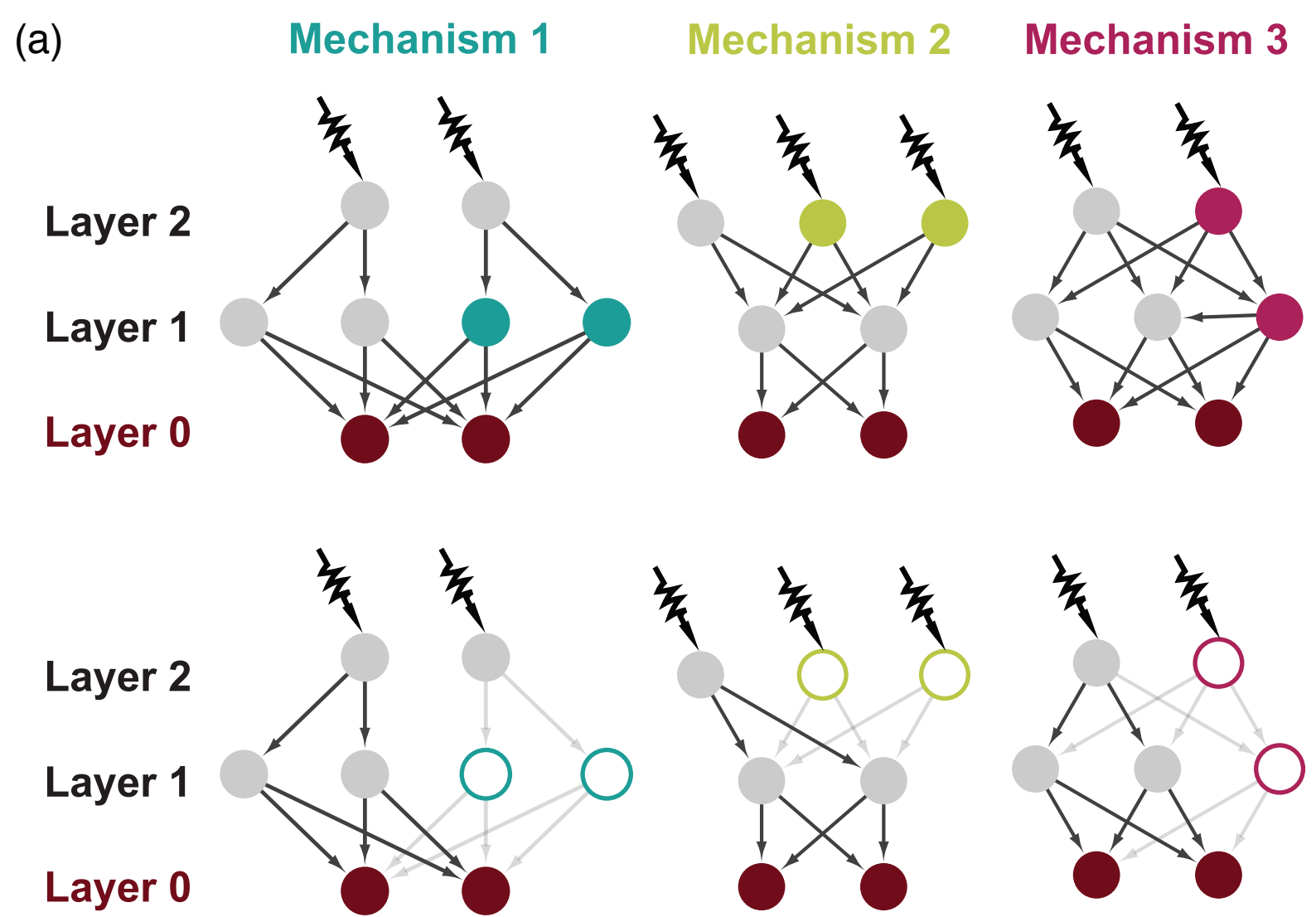

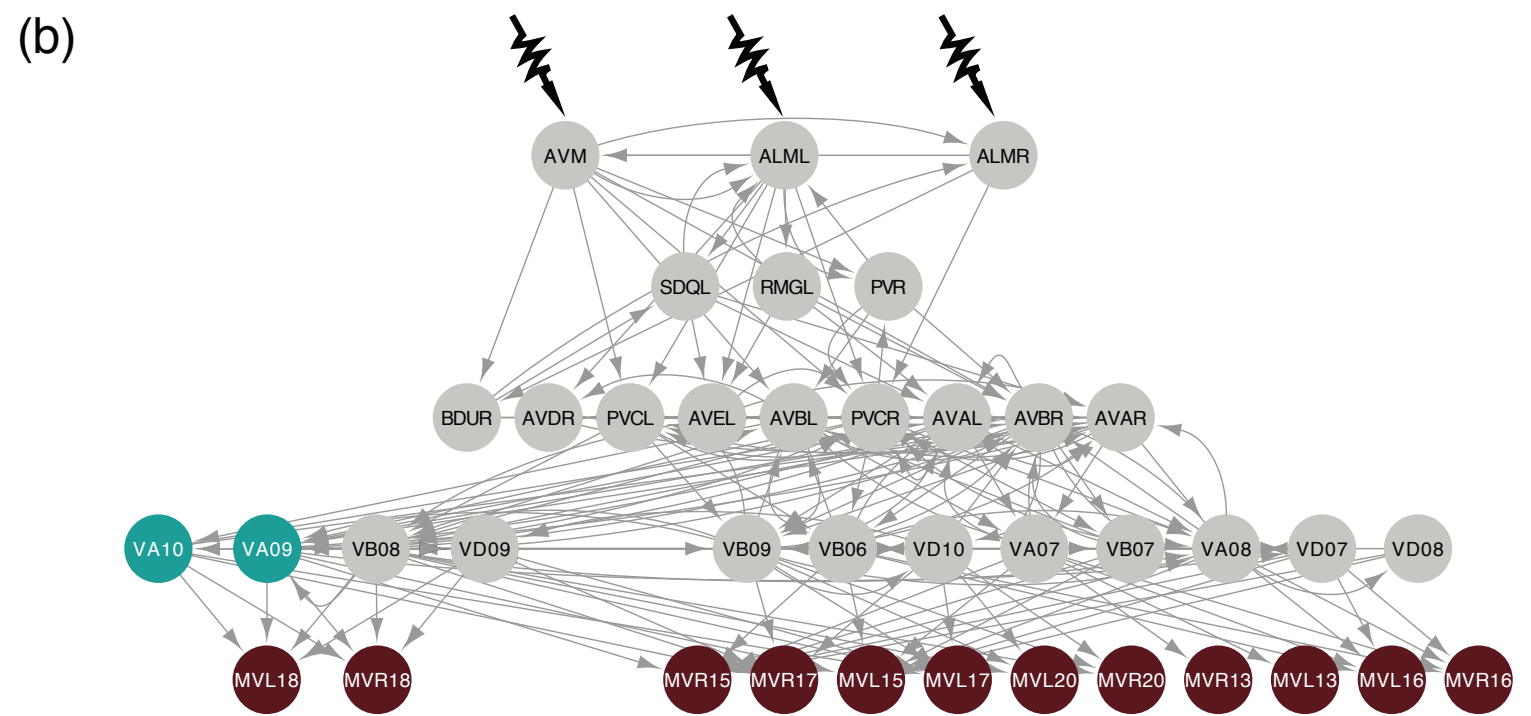

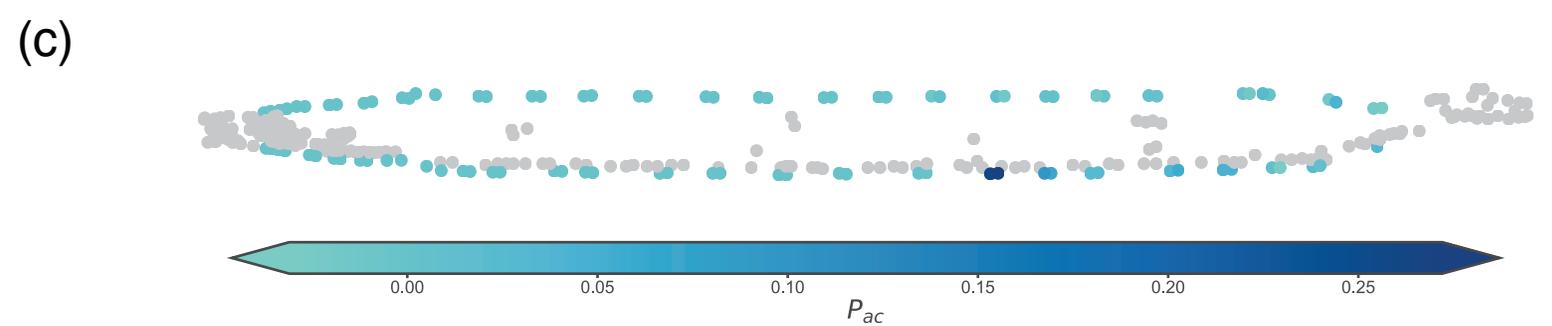

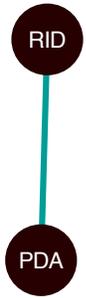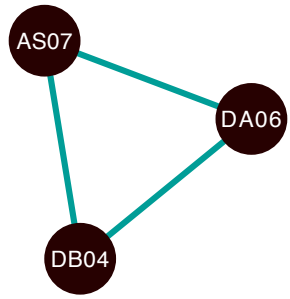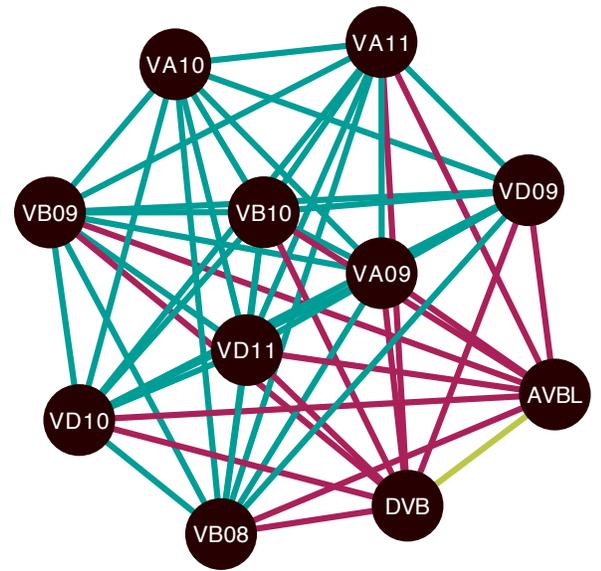

**Mechanism 1**  **Mechanism 2**  **Mechanism 3**

**Figure 2: Control mechanisms behind the synthetic essentiality of neuronal pairs.** (a) Three distinct mechanisms are observed for the reduction in control found in synthetic essential pairs. 16 pairs are predicted to lose fine muscle control via Mechanism 1, one by Mechanism 2, and 40 by Mechanism 3. Mechanism 1: (Upper) Since there are more Layer 1 neurons receiving independent control signals than there are muscles, the muscles are controllable. (Lower) The removal of one Layer 1 and one Layer 2 neuron causes there to be only one independent control signal arriving at the two neurons in Layer 1, and consequently only one independent control signal arriving at the muscles. Mechanism 2: (Upper) Since there are the same number of Layer 1 neurons with independent control signals as there are muscles, the muscles are controllable. (Lower) The removal of two Layer 2 neurons leaves only one independent control signal arriving at Layer 1, as in Mechanism 1. Only one muscle may be independently controlled. Mechanism 3: (Upper) Since there are more Layer 1 neurons receiving independent control signals than there are muscles, the muscles are controllable. (Lower) The removal of two Layer 1 neurons disconnects the pathway from one of the Layer 2 neurons to the muscles. Again, only one control independent control arrives at Layer 1, and thus also the muscles. (b) Example synthetic essential pair, VA09 and VA10. This double ablation is predicted to affect control of the two muscles MVR18 and MVL18 via Mechanism 1. (c) The probability of ablation-induced loss of control over each muscle in the body of *C. elegans* following ablation of the pair {VA09, VA10}. Muscles most likely to lose control are coloured dark blue, and the least likely to lose control in green. The location of neuron cell bodies are shown in grey.

**Figure 3: Neurons involved in synthetic essentiality via double ablations.** The neurons comprising pairs of synthetic essential neurons are shown as networks, in which two neurons are linked if they occur together in a synthetic essential pair. Edge colour describes the Mechanism of essentiality.

We identified only 57 synthetic essential pairs, which we predict to lead to loss of control via one of three distinct mechanisms. These mechanisms are rooted one layer up from the muscles, acting to reduce the number of independent control signals received by the motor neurons (and then consequently the muscles). Many of the neurons involved are ventral cord motorneurons, and the synthetic essential pairs predict their functional interdependence.

*Synthetic triplets.*

Of all the 3,580,779 possible triple ablations, 2,847,827 (79.5%) are predicted to have no effect and 732,952 are predicted to lead to a loss of control. 731,525 of these essential triplets are explained through the simple summation of effects of essential individual neurons and synthetic essential pairs - see Table 2 and Supplementary Tables S2 and S5. Only the remaining 46 triplets, and 1381 negative interactions, exhibit synthetic essentiality, a remarkably small number. These synthetic essential triplets comprise 28 distinct neurons and are grouped into four groups based on the set of muscle cells they impact (which comprise one, two, four, or eight muscles). Again, we find that the number of independently controllable muscles reduces by a maximum of one for each ablated neuron, and hence we observe a reduction of no more than three (leaving 86 controllable muscles) following a triple ablation. We observe negative interactions, i.e. a greater loss of control than expected by the simple summation of the loss due to the single and double ablations, in 1381 cases (see SI Section II); 1363 triplets contain only one essential neuron, yet lead to a reduction in control of two muscle cells, and 18 contain two essential neurons and lead to a reduction in control of three muscle cells.

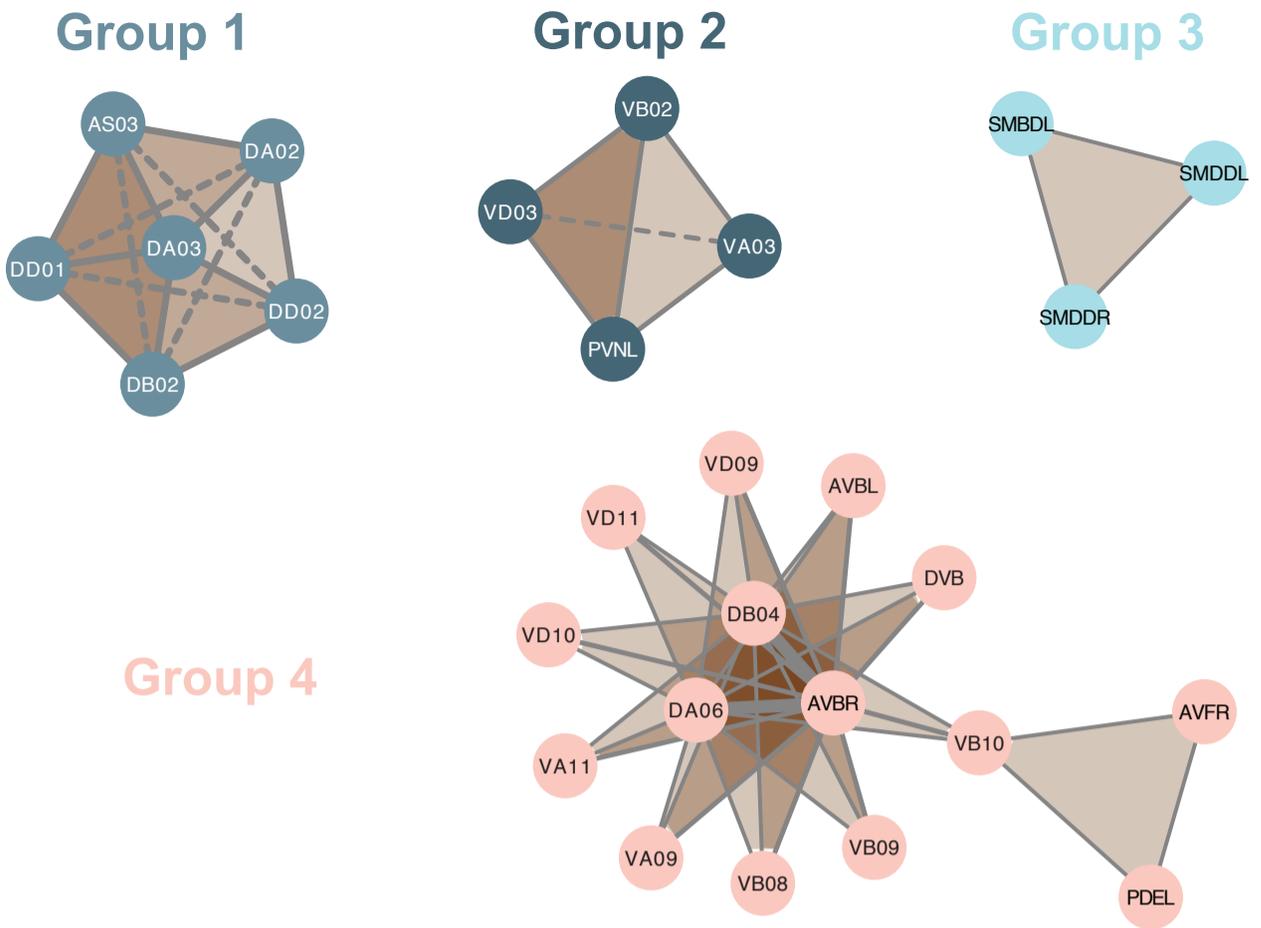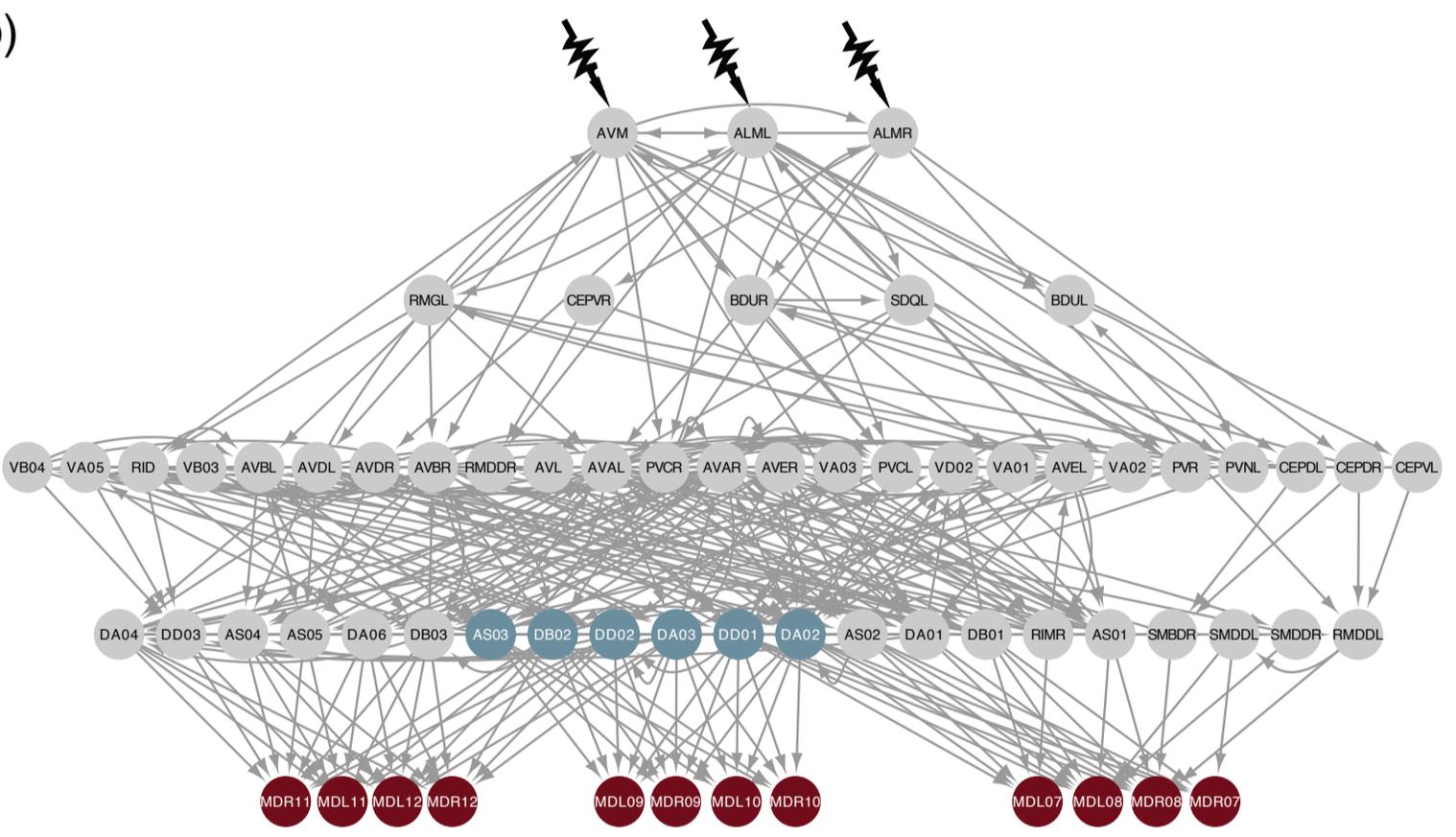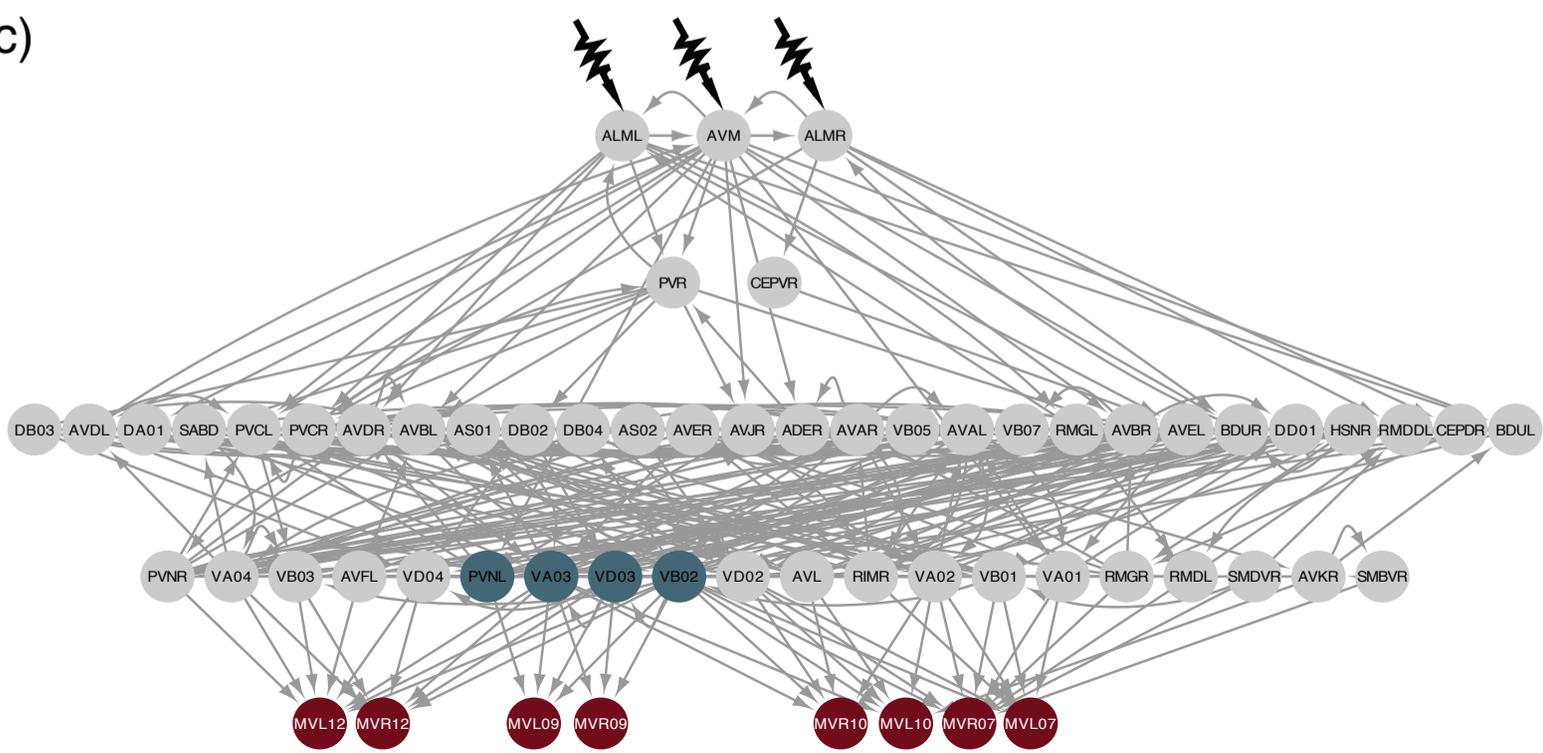

**Figure 4: Groups of neurons involved in synthetic essentiality via triple ablations.** (a) The neurons comprising triplets of synthetic essential neurons can be divided into four groups, of sizes six, four, three, and 15 neurons respectively. In the depicted networks, two neurons are linked if they occur together in a synthetic essential triplet, and edge width increases with the number of such triplets. Each entire triplet of three is a simplicial complex, shown as a shaded triangular face. In Groups 1-3, any selection of three neurons from the group constitute a predicted synthetic essential triplet. This is not the case for Group 4, which comprises 21 triplets which have at least one neuron in common with each other. (b) Group 1 and the relevant control pathways. The set of muscle cells {MDL09, MDR09, MDL10, MDR10} connect to the nervous system via neuromuscular junctions to only the six neurons in Group 1. To control each of these muscles independently requires four independent controls, one for each muscle cell. Thus the removal of any three of the neurons in Group 1 leads to at most three independent control signals arriving at the muscles, and a maximum of three independently controllable muscles. (c) Group 2 and the relevant control pathways. Similarly to the mechanism (b), the ablation of any three of the neurons {PVNL, VA03, VD03, VB02} lead to a loss of independent control over the muscle set {MVL09, MVR09}.

*Mechanisms.*

We identified four distinct groups of neurons within the synthetic essential triplet predictions – see Figure 4 and Supplementary Figure S1. Within Groups 1 and 2, one of six neurons and one of four, any combination of three neurons from the set leads to a predicted loss of control. The abstracted Group networks of triplets are *simplicial complexes* (28,29) (Figure 4(a)). In networks, nodes are connected pairwise via links which represent dyadic relationships. Yet, polyadic relationships, in which three or more nodes interact simultaneously, abound in real systems. In algebraic topology, these generalised connections are called simplexes, and the system as a whole, a simplicial complex. Triplets can be described as 2-simplexes, and the essential pairs in Figure 3 as 1-simplexes. To understand the mechanistic origins of these triplets, we turn back to the network. In each case, the set of $N$ neurons directly connects to $N-2$ muscles – which have no other connections to other neurons. Thus by ablating any choice of three neurons from the set, a maximum number of $N-3$ independent control signals can reach the $N-2$ muscles, and we lose controllability of one of them. For example, the four muscle cells {MDL09, MDR09, MDL10, MDR10} are connected to the nervous system via six neurons – see Figure 4(b). If we remove, say, AS03, DA02, and DA03, only three neurons provide the path to the muscles. This means that only three control signals can reach four muscles, and they can no longer each be independently controlled. Group 3, comprising three neurons, leads to a trivial disconnection of a muscle cell. Group 4 represents an overlap with synthetic essential pairs – see below.

Group 1: (20 synthetic essential triplets). Any selection of three neurons from the set of six {AS03, DA02, DA03, DB02, DD01, DD02} will reduce the control over a group of four muscle cells, {MDL09, MDR09, MDL10, MDR10}. Given that these six neurons provide the only connections between these four muscles and the nervous system, a maximum of six independent control signals can reach them. When three are ablated, this number is reduced to three independent control signals, one less than the number of muscle cells.

Group 2: (4 synthetic essential triplets). Any selection of three neurons from the group of four {PVNL, VA03, VB02, VD03} leads to a predicted reduction in fine control over two muscle cells, {MVL09, MVR09}; see Figure 4(c).

Group 3: (1 synthetic essential triplet). When the group of three neurons {SMBDL, SMDDL, SMDDR} are ablated, the muscle MDR06 is disconnected entirely from the network. See SI Section II.

Group 4: (21 synthetic essential triplets). The 15 unique neurons in this group overlap with the synthetic essential pairs in terms of individual components, but never including a complete pair. They also impact the same muscle cells, with eight of the ten posterior cells affected by the synthetic essential pairs predicted to be affected. Unlike the first three groups, not any combination of three will suffice, yet clear patterns emerge: 20 of the 21 triplets contain AVBR, and either DA06 or DB04, and are predicted to impact a small set of dorsal muscles in the mid-section of the worm. These triplets are then completed by one of 10 individual neurons (see Supplementary Table S5) – any of the 10 with each of the pairs {AVBR, DA06} or {AVBR, DB04}. The final prediction {VB10, PDEL, AVFR} links to the rest of the group only via one of the 10 individual neurons (VB10).

Groups 1-3, which comprise only neurons not involved at all in single or double ablations, uncover small groups of muscles with only few connections to the nervous system (indeed, only via these precise neurons), which are therefore susceptible to loss of control following damage to these Groups. Given the repeating connectivity patterns observed in the motor system, this is perhaps not surprising (30,31), and increasing numbers of ablations might expect to uncover further units. Group 4 contains neurons which are also involved in synthetic essential pairs, and centres on the command interneuron class AVB and a number of posterior dorsal and ventral motorneurons. Their appearance together suggests a more complex functional interdependence.

**Discussion**

The most striking aspect of the synthetic essential neuronal pairs and triplets predicted above is their level of specificity. Our methodology identifies highly localised groups of neurons and affected muscles, with consistent network level mechanisms of loss of control. The small size of these subsets, comparable to that of single ablations, is in contrast to studies in the genetic realm which, unsurprisingly given the combinatorics, find many orders of magnitude more essential digenic and trigenic interactions than singly essential genes. Indeed, the trigenic network tends to be ~100 times larger than the digenic network – itself containing ~1000 times more essential combinations than there are singly essential genes. Further, we observe only very few cases of negative interactions and no positive interactions, the dominant results in genetic interactions (15,16). About 83% of budding yeast genes are non-essential, whereas the control framework predicts that about 93% of neurons are non-essential. Double and triple knockout studies have found that 2.3% of double and 1.6% of triple genetic interactions are negative. We predict that only 0.0003% of double neuronal ablations lead to a negative interaction, and 0.04% of triple ablations. The effects of genetic interactions are found to be weaker in magnitude with the addition of more knockouts, yet the phenotypes predicted following synthetic essential neuronal ablations will be similarly different to the healthy worm, given the loss of control over one muscle cell. However, it is not clear if these differences are significant or meaningful. The analysis we present here is based on structural control, which provides discrete results rather than the continuous measurements accompanying genetic knockouts. This translates to a binary decision – control is lost or not – and limited measurements of relative strength. Structural control also does not allow for the possibility of positive interactions: a second or third ablation will not increase muscle controllability.

Yet, understanding and cataloguing the effects of multiple ablations can inform *rescue.* In genetics, the "Lazarus effect" is the counterintuitive restoration of lost function through gene *deletion* (32). In neural systems, this translates to the possibility of restoring function via neuronal ablation. If neuronal damage has led to the loss of control of a function, such as locomotion, we can seek to restore the function by removing one or more further neurons. According to structural control, we cannot recover control through the removal of neurons and/or links. However, if the extension is made to a more sophisticated analysis incorporating link weights and energetic considerations (8,9), we might be able to find such synthetic rescue combinations. Indeed, it is theoretically possible to reduce the energy required to reach certain configurations of the behavioural state space, i.e. desired locomotory patterns, by removing neurons and/or links, and thus potentially rescuing lost function.

The predictions for synthetic essential neuronal pairs and triplets also begin to reveal the functional dependence of particular muscle groups on small sets of neurons. The double ablations show how damage to the network upstream of the muscles (up to Layer 0) can lead to loss of control downstream. The removal of particular pairs reduces the number of independent control signals arriving at Layer 1, which goes on to have an effect on control of the muscles themselves. For the triple ablations, small groups of muscles exist which only receive control signals from a small number of neurons. Ablating fractions of these sets of neurons leads to fewer signals arriving at the muscle cells, and a loss of controllability. Note that it is often not important *which* three neurons in these sets, just the total of three. This highlights an amount of redundancy, in the control sense, in the network, which provides robustness to failure – it is not until at least three neurons are removed that the effect is felt. Further, each additional ablated neuron corresponds to a maximum reduction in controllability of one muscle cell.

Finally, the unexpectedly small and precise sets of falsifiable predictions lend themselves well to experimental confirmation. The identification of a small number of muscle cells will allow for the specific characterisation of the predicted phenotype, to reduce the space of hundreds of locomotor features that tracking software can measure. Cell-specific laser ablation, or advances in genetic ablation techniques allowing one to distinguish between neurons in a class, can be employed. The experiments will still face two major challenges: (i) Making the appropriate strains. Each class of neuron will need a different marker line, so the triple ablation predictions which cover three different classes will need at a minimum three different labeling transgenes with three different colors; (ii) Total number of ablations. To properly control the experiments, a mock-ablation would need to be performed for each reporter combination, and to interrogate the predictions for triple ablations, a number of ablation and mock-ablation experiments will need to be performed on combinations of two neurons within the set of three. Finally, we note that a number of the synthetic essential pairs unfortunately contain at least one neuron that lies within a region of the network with only partial (11 pairs) or missing data (19 pairs), and knowledge of network structure is supplemented by inference (31). Thus further improvements to the wiring diagram are warranted prior to experimental testing these particular pairs. The remaining 27 pairs lie in regions with more reliable connectivity maps.

**Materials and Methods**

*The C. elegans wiring diagram: data.*

We base our analyses upon the mapping of the *C. elegans* connectome presented in (33). This wiring diagram comprises 279 nonpharnygeal neurons connected by 2,194 directed synaptic connections and 1,028 reciprocal gap junctions. 95 muscles connect to the nervous system via 552 neuromuscular junctions to 124 motor neurons. See SI Section V for analyses and discussion concerning recent updates to the connectome (34).

*Structural controllability.*

Following the approach in (3), we model the nematode nervous system as a directed network whose nodes include neurons and muscles, and whose links represent the electrical and chemical synaptic connections between them, including neuromuscular junctions. Formally, the dynamics of the system composed of *N* neurons and *M* muscles is described by

$$\dot{z}(t) = f(z, v, t), \qquad (1)$$

where $z(t) = [z_1(t), z_2(t), \ldots, z_{N+M}(t)]^\mathsf{T}$ denotes the states of *N+M* nodes at time *t*, $f(*) = [f_1(*), f_2(*), \ldots, f_{N+M}(*)]^\mathsf{T}$ captures the nonlinear dynamics of each node, and $v(t) = [v_1(t), v_2(t), \ldots, v_S(t)]^\mathsf{T}$ represents the external stimuli applied to the *S* touch receptor neurons. Assuming that in the absence of additional stimuli the nervous system is at a fixed point $z^*$, where $f(z^*, v^*, t) = 0$, and using $x(t) = z(t) - z^*$ and $u(t) = v(t) - v^*$, Eq. (1) can be linearised,

obtaining

$$\begin{cases} \dot{x}(t) = Ax(t) + Bu(t), \\ y(t) = Cx(t), \end{cases} \quad (2)$$

where $A \equiv \frac{\partial f}{\partial z}|_{z^*, v^*}$ corresponds to the adjacency matrix of the connectome, with non-zero elements $A_{ii}$ that represent the nodal dynamics of node $i$; the input matrix $B \equiv \frac{\partial f}{\partial v}|_{z^*, v^*}$ represents the receptor neurons on which the external signals are imposed, e.g. ALML/R and AVM for anterior gentle touch; and the vector $y(t)$, selected by the output matrix $C$, represents the states of the *M* muscle cells. In other words, the response of *C. elegans* to external stimuli can be formalised as a target control problem (26), asking if the inputs received by receptors in $B$ can control the state of the muscles listed in $C$. The muscles are controllable if, with a suitable choice of inputs $u(t)$, they can move in any desired manner, i.e. $y(t)$ can reach an arbitrary position of the *M*-dimensional state space (35). To determine this, we consider the controllability matrix, given by $K = [CB, CAB, CA^2B, \dots, CA^{N+M-1}B]$. Kalman's criterion (36) tells us that the system (2) is deemed structurally controllable if $rank\ K = M$; this translates to the situation that all *M* muscles are controllable via signals from the input sensory neurons. Moreover, $rank\ K$ is equal to the number of controllable muscles, thus providing a way to measure the level of controllability (3).

**Acknowledgements**

Funding for E.K.T. and A.-L.B. was provided by the NSF under award no. 1734821.

**Author Contributions**

E.K.T. and A.-L.B. designed the research. E.K.T. performed the research. E.K.T. and A.-L.B. wrote the paper.

**Data availability statement**

The *C. elegans* wiring diagram is publicly available online (33,34), and details of all results are either provided as Extended Data or can be found at https://github.com/EmmaTowlson/c-elegans-control.

**Code availability statement**

All analyses were completed in Python using scripts available at https://github.com/EmmaTowlson/c-elegans-control.

**Competing Interests**

The authors declare no competing financial interests.

**Correspondence**

Correspondence should be addressed to a.barabasi@northeastern.edu.

# Tables

|  | Reduction | No effect |
|---|---|---|
| $E_{single}E_{single}$ | 190 | 0 |
| $E_{single}N_{single}$ | 5180 | 0 |
| $N_{single}N_{single}$ | **57** | 33353 |

**Table 1**: **Double ablation predictions and relation to single ablation predictions.** A selected pair of neurons may comprise two individually essential neurons ($E_{single}$), one essential and one non-essential neuron ($N_{single}$), or two non-essential neurons. Synthetic essential pairs with no overlap with single ablation predictions are coloured **red**. Totals exclude the trivial case of removal of all input neurons.

|  |  | Reduction | No effect |
|---|---|---|---|
| Pairs | $3E_{pair}$ | 149 | 0 |
|  | $2E_{pair}$ | 16 | 0 |
|  | $E_{pair}E_{single}$ | 44 | 0 |
|  | $E_{pair}N_{single}$ | 14170 | 0 |
| No pairs | $E_{single}E_{single}E_{single}$ | 1140 | 0 |
|  | $E_{single}E_{single}N_{single}$ | 49210 | 0 |
|  | $E_{single}N_{single}N_{single}$ | 668176 | 0 |
|  | $N_{single}N_{single}N_{single}$ | **46** | 2847827 |

**Table 2**: **Triple ablation predictions and relation to double and single ablation predictions.** A selected triplet of neurons is comprised of combinations of individually essential neurons ($E_{single}$), individually non-essential neurons ($N_{single}$), and synthetic essential pairs of neurons ($E_{pair}$). Synthetic essential triplets with no overlap with single ablation predictions, and not containing a complete pair of synthetic essential neurons, are coloured **red**. Totals exclude the trivial case of removal of all input neurons.

# Synthetic ablations in the *C. elegans* nervous system

Supplementary Information

## I. Detailed results: Double ablations

As discussed in the main text, we recover 57 synthetic essential pairs, with only three Mechanisms of loss of control. We also identify a small number of double ablations which, while expected to result in a loss of control due to the involvement of one or more essential single neurons, do not result in the precise level of controllability loss expected from the essential single neuron(s) alone; see Tables S1 and S4, and Extended Data 1. Most notably, we find only one example of enhancement of control loss. A pair comprising one singly essential neuron and one non-essential neuron ($E_{single}N_{single}$) would be expected to result in the loss of fine control over one muscle cell, due to the essential neuron. We find one such pair where the value of loss of control increases to two muscle cells: {AS11, RID}. Here AS11 is essential alone, and RID is non-essential.

We also note four instances where the loss of control is diminished from the expectation. We would expect a pair comprising two singly essential neurons ($E_{single}E_{single}$) to result in the loss of fine control over two muscle cells if we were to simply sum the effects – one muscle for each ablated neuron. We recover four such pairs which only lead to a loss of control over *one* muscle cell: {AVAL, AS08}, {AVAR, AS08}, {AVAL, DA07}, and {AVAR, DA07}. This can be explained by an overlap in network neighbourhood; the same muscle cells are affected by the ablation of each individual neuron, so the impact of their simultaneous removal is contained to the same cells.

**II. Detailed results: Triple ablations**

We identify four groups of synthetic essential triplets. The network effects behind Groups 1 and 2 are shown in Figure 4(b) and (c). When the three neurons in Group 3 {SMBDL, SMDDL, SMDDR} are ablated, the muscle MDR06 is disconnected entirely from the network – see Figure S1. We do not show Group 4 as its constituents are not localised in the network. In addition to these groups, we find a number of predictions for enhancement of control loss. These triplets contain singly essential or synthetic essential pairs, but are predicted to lead to a greater reduction in controllability than expected from the summation of effects alone; see Tables S2 and S5. Specifically:

(i) A triplet comprising a synthetic essential pair and an essential single neuron ($E_{pair}E_{single}$) would be expected to result in the loss of fine control over two muscle cells in the case of a simple summation of effects – one for the pair, and one for single neuron. We identify one such triplet, {PDA, AS11, RID}, which results in the loss of control over *three* muscle cells. This is consistent with the observation that {PDA,RID} is a synthetic essential pair, and as noted above the ablation of {AS11,RID} leads to a loss of control over two muscle cells.

(ii) A triplet comprising two singly essential neurons and one non-essential neuron ($E_{single}E_{single}N_{single}$) would also be expected to result in the loss of fine control over two muscle cells, one for each essential neuron. We identify 18 triplets where this loss of control is increased to three muscle cells. All of these triplets contain the pair {AS11, RID} and one further neuron (see Extended Data 2), again consistent with the enhancement of loss of control found in the double ablations.

(iii) A triplet comprising one single essential neuron and two non-essential neurons ($E_{single}N_{single}N_{single}$) would be expected to lead to a loss of control in one muscle cell. We find 1363 triplets where this loss is predicted to be increased to two muscle cells (see Extended Data 2).

Finally, we predict a number of triple ablations will result in a smaller reduction in controllability than expected by the summation of the effects of parts. This is due to an overlap in the neurons and muscles involved in each case. In summary, these cases are: 149 triplets of the form $3E_{pair}$; 16 of the form $2E_{pair}$; 43 of the form $E_{pair}E_{single}$; 68 of the form $E_{single}E_{single}E_{single}$; and 1036 of the form $E_{single}E_{single}N_{single}$. See also Table S2.

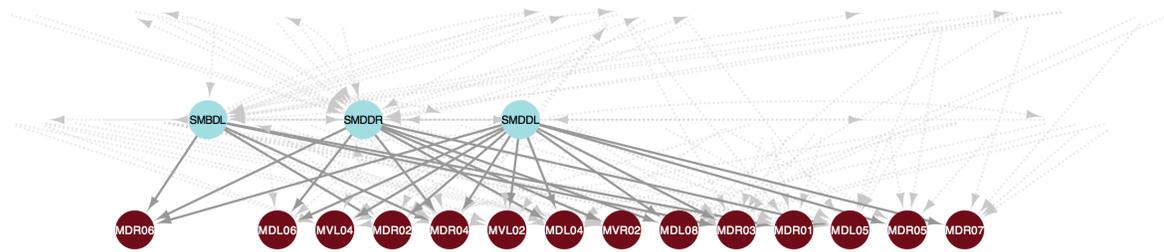

**Figure S1**: Group 3. The removal of {SMBDL, SMDDL, SMDDR} disconnects the muscle cell MDR06.

## III. Affected muscles

The structural controllability framework provides a deterministic number of controllable muscles, but there are in general multiple configurations of muscles that could comprise these controllable cells (2). Therefore, while we find 89 independently controllable muscles in the healthy worm, the precise set of 89 muscles is not unique: there are multiple solutions to the control problem (7), each of which give rise to the same level of controllability. By cataloguing these independent solutions, we can assign a probability to which muscles are more likely to experience a reduction in control. Specifically, for each ablation of pairs or triplets of neurons, we numerically obtained the probability pattern of each muscle losing its controllability, and compared this pattern to that of the healthy worm. We obtained these patterns through 1000 iterations of the structural controllability analysis. The difference between the two probability patterns reveals which muscles are affected most strongly by the ablation. The muscle patterns tend to be highly spatially co-localised, offering quite specific predictions pertaining to expected phenotypes in future experiments.

Indeed, the synthetic essential pairs are predicted to affect only a small set of muscles in the ventral posterior section of the body. Figure S2 shows three exemplary pairs, and the consistent region predicted to experience a loss of control. Probability patterns for all predictions can be found at https://github.com/EmmaTowlson/c-elegans-control.

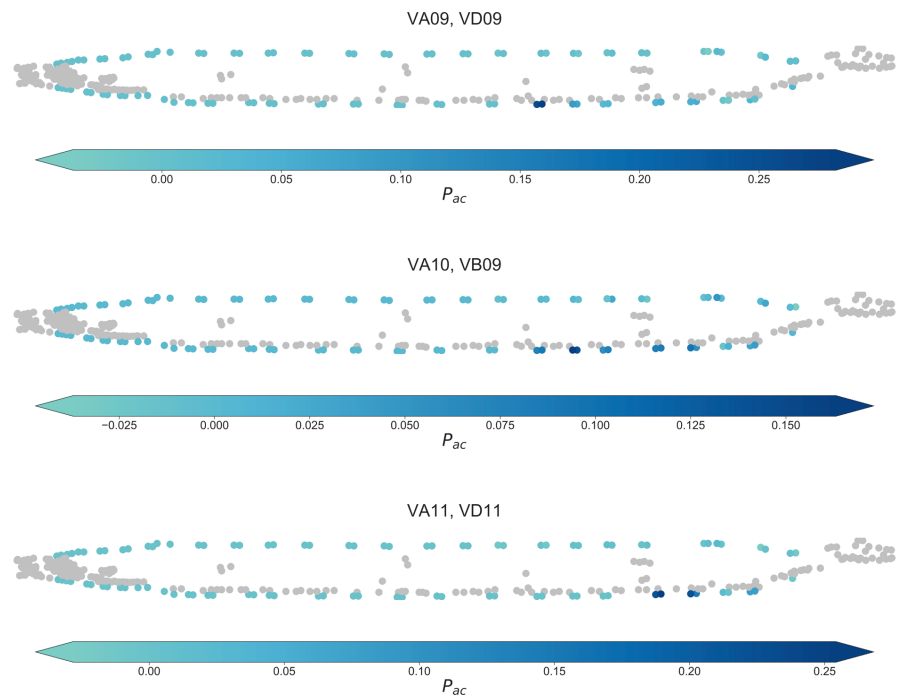

*Figure S2*: *Affected muscles for synthetic essential pairs.* The probability of ablation-induced loss of control over each muscle is shown in the body of *C. elegans* following ablation of the pair {VA09, VD09} (top), {VA10, VB09} (middle), and {VA11, VD11} (bottom). Muscles most likely to lose control are coloured dark blue, and the least likely to lose control in green. The locations of neuron cell bodies are shown in grey. A similar area in the ventral posterior section of the body is affected most strongly in each case.

**IV: Alternative input neurons**

The analysis conducted in the main paper considers the locomotory response to gentle touch on the anterior and posterior body, which translates to the input neurons in the vector $B$ in Equation 2, being {ALML, ALMR, AVM} and {PLML, PLMR} respectively. We considered three further sets of sensory neurons for the input set, each known to elicit a locomotory response upon stimulation: {FLPL, FLPR}, {PVDL, PVDR}, and {ASHL, ASHR}. In each case, we ablated *in silico* the 58 synthetic essential pairs and 1327 synthetic triplets that we uncovered in our original analysis as resulting in a greater loss of controllability than expected by the summation and/or overlap of lower order results. We recovered precisely the same findings in terms of quantifying the loss of controllability, suggesting, as per our original study (6), that these control properties are more general to locomotion in *C. elegans* rather than specific to just one behaviour. In other words, the uncovered synthetic pairs and triplets are not just important for the response to gentle touch, but also for other locomotion-based behaviours.

**V. The *C. elegans* wiring diagram: alternative data**

The analyses in the main text are based upon the mapping of the *C. elegans* connectome presented in (1). The imperfections in the wiring diagram have recently come under scrutiny from the community (2–4), and indeed the original EM images were recently reexamined, resulting in a new wiring diagram with significant differences in connectivity (5). As such, we reexamined our main findings on this version. The new connectome has some significant differences which must be addressed first:

(i) Electrical coupling between the body wall muscle cells is modelled in the form of bidirectional edges between neighbouring muscle cells. This coupling is not modelled in the wiring diagram used in the main text. It is important to note that the structural control framework relies on a path-based approach, and incorporating electrical coupling in this manner simply leads to the conclusion that all muscle cells are always controllable: when a control signal reaches any muscle cell, it can continue along the paths between the muscles themselves indefinitely. There is no current methodology to appropriately account for the difference in the properties of different connection types (2), so we removed these coupling links.

(ii) This dataset is enriched with the connectivity of numerous end organs not present in the earlier wiring diagram. To facilitate a direct comparison with our main results, and to remain focused on control of locomotion as driven by the body wall muscles, we removed the extra end organs. This ensured the same set of 95 muscle cells as the target nodes in the structural control problem.

(iii) Finally, there are a significant number of extra links in the updated network. At this high density, we recover fewer results from the control framework (see Table S6). Any inevitable mapping errors are more likely to lie within the weakest connections, and they are more likely to be sites of variation between individuals. Therefore it is reasonable to hypothesise that a robust control structure should exist among the stronger, or more reliable, links. We systematically pruned the weakest links, i.e. those with the fewest synapses between neuron pairs, from the network. We considered thresholds of $\tau = 0 - 3$ neurons as criteria for the presence or absence of an unweighted link. For $\tau = 0$ all links remain, and for $\tau = 3$ links with fewer than 3 synapses were removed.

These processing steps provided us with a connectome with $N + M = 375$ nodes (the 374 neurons and muscles in the original connectome, plus VC06, a neuron which is disconnected in the previous dataset) and $L = 6022$ directed unweighted links. With an increasing threshold this reduced the number of links to 4124 ($\tau = 1$), 3068 ($\tau = 2$), and 2318 ($\tau = 3$).

Firstly, we examined whether or not our original predictions for single neuron ablations were robust to the differences in the two wiring diagrams; our findings can be found in Table S6. We recover the most complete set of predictions at a threshold of three synapses, i.e. where links with only one or two synapses are pruned from the network. At this threshold, we find 91 controllable muscles in the healthy worm, and recover all predictions from the original analysis (6) plus an extra twelve neurons. These twelve neurons comprise eleven ventral motor neurons plus PDA, and offer potential further single-cell predictions for experimental testing.

We focused our next analyses on the connectome with the greatest correspondence with our original predictions – that with the threshold of 3 synapses. We repeated the structural control analysis for the 57 synthetic essential double ablations and the 46 synthetic essential triple ablations. We observe large agreement with the synthetic essential pairs: 53 also lead to a reduction in controllability in this connectome, while the final 4 do not. We find less consistency when examining the synthetic essential triplets: only 17 of the 46 are predicted to reduce control. While the two connectomes initially recover similar results for the control analysis, as we move to higher order interactions we encounter more differences. Given the large differences between the two wiring diagrams this is not surprising, as the finer details of the network structure become more important with the removal of more neurons. More accurate and modern maps will improve the accuracy and completeness of the predictions higher order interactions. Nevertheless, these results are encouraging. The consistency in the double ablation predictions in particular highlights a large degree of robustness of the control organisation to rewired links.

## Supplementary tables

|  | Reduction (-2) | Reduction (-1) | No effect |
|---|---|---|---|
| $E_{single}E_{single}$ | 186 | 4 | 0 |
| $E_{single}N_{single}$ | **1** | 5179 | 0 |
| $N_{single}N_{single}$ | 0 | **57** | 33353 |

**Table S1**: **Double ablation predictions and amount of reduction in control.** As per Table 1, a selected pair of neurons may comprise two individually essential neurons ($E_{single}$), one essential and one non-essential neuron ($N_{single}$), or two non-essential neurons. A loss of control over one (-1) or two (-2) muscles is predicted from double ablations. Synthetic essential pairs with no overlap with single ablation predictions are coloured **red**. One pair is predicted to lead to a greater loss of control than explained by the presence of a single essential neuron alone, coloured in **blue**. Predictions commensurate with the simple summation and/or overlap of effects of single ablations are coloured black. Totals exclude the trivial case of removal of all input neurons.

|  |  | Reduction (-3) | Reduction (-2) | Reduction (-1) | No effect |
|---|---|---|---|---|---|
| Pairs | $3E_{pair}$ | 0 | 149 | 0 | 0 |
|  | $2E_{pair}$ | 0 | 0 | 16 | 0 |
|  | $E_{pair}E_{single}$ | **1** | 0 | 43 | 0 |
|  | $E_{pair}N_{single}$ | 0 | 0 | 14170 | 0 |
| No pairs | $E_{single}E_{single}E_{single}$ | 1072 | 68 | 0 | 0 |
|  | $E_{single}E_{single}N_{single}$ | **18** | 48156 | 1036 | 0 |
|  | $E_{single}N_{single}N_{single}$ – no P | 0 | **1363** | 666813 | 0 |
|  | $N_{single}N_{single}N_{single}$ – no P | 0 | 0 | **46** | 2847827 |

**Table S2**: **Triple ablation predictions and amount of reduction in control.** As per Table 2, a selected triplet of neurons is comprised of combinations of individually essential neurons ($E_{single}$), individually non-essential neurons ($N_{single}$), and synthetic essential pairs of neurons ($E_{pair}$). A loss of control over one (-1), two (-2), or three (-3) muscles is predicted from triple ablations. Synthetic essential triplets with no overlap with single ablation predictions, and not containing a complete pair of synthetic essential neurons, are coloured **red**. Triplets predicted to lead to a greater loss of control than explained by the presence of one or more essential neurons or neuron pairs alone are coloured in **blue**. Predictions commensurate with the simple summation and/or overlap of effects of single and/or double ablations are coloured black. Totals exclude the trivial case of removal of all input neurons.

| |
|---|
| AS08 |
| AS09 |
| AS10 |
| AS11 |
| AVAL |
| AVAR |
| DA07 |
| DA08 |
| DA09 |
| DB05 |
| DB06 |
| DB07 |
| *DD04* |
| *DD05* |
| DD06 |
| *PDB* |
| VA12 |
| VB11 |
| VD12 |
| VD13 |

**Table S3**: **Neuronal components of the single ablation predictions.** Those in *italics* have been experimentally verified (1). All ablations are predicted to lead to the loss of independent control over one muscle cell.

| Synthetic essential pair | | Mechanism |
|---|---|---|
| AS07 | DB04 | |
| AS07 | DA06 | |
| DA06 | DB04 | |
| PDA | RID | |
| VA09 | VB09 | |
| VA09 | VD09 | |
| VA09 | VB08 | Mechanism 1 |
| VA09 | VB10 | |
| VA09 | VD11 | |
| VA09 | VD10 | |
| VA09 | VA10 | |
| VA09 | VA11 | |
| VA10 | VB09 | |
| VA10 | VD09 | |
| VA10 | VB08 | |
| VA10 | VB10 | |
| VA10 | VD11 | |
| VA10 | VD10 | |
| VA10 | VA11 | |
| VA11 | VB09 | |
| VA11 | VD09 | |
| VA11 | VB08 | |
| VA11 | VB10 | |
| VA11 | VD11 | |
| VA11 | VD10 | |
| VB08 | VB09 | |
| VB08 | VD09 | |
| VB08 | VB10 | |
| VB08 | VD11 | |
| VB08 | VD10 | |
| VB09 | VD09 | |
| VB09 | VB10 | |
| VB09 | VD11 | |
| VB09 | VD10 | |
| VB10 | VD09 | |
| VB10 | VD11 | |
| VB10 | VD10 | |
| VD09 | VD11 | |
| VD09 | VD10 | |
| VD10 | VD11 | |
| AVBL | DVB | Mechanism 2 |
| AVBL | VB09 | |
| AVBL | VA09 | |
| AVBL | VD09 | |
| AVBL | VB08 | |
| AVBL | VB10 | |
| AVBL | VD11 | Mechanism 3 |
| AVBL | VD10 | |
| AVBL | VA11 | |
| DVB | VB09 | |
| DVB | VA09 | |
| DVB | VD09 | |

| | |
|---|---|
| DVB | VB08 |
| DVB | VB10 |
| DVB | VD11 |
| DVB | VD10 |
| DVB | VA11 |

**Table S4**: **Neuronal components of the double ablation predictions**. Synthetic essential pairs arise from three mechanisms (see Figure 2(a) and Figure 3).

| Synthetic essential triplet | | | Group |
|---|---|---|---|
| AS03 | DA03 | DD02 | Group 1 |
| AS03 | DA02 | DD01 | |
| AS03 | DA03 | DD01 | |
| AS03 | DA03 | DA02 | |
| AS03 | DA02 | DD02 | |
| AS03 | DD01 | DD02 | |
| DA02 | DD01 | DD02 | |
| DA03 | DA02 | DD02 | |
| DA03 | DA02 | DD01 | |
| DA03 | DD01 | DD02 | |
| DB02 | AS03 | DD02 | |
| DB02 | DA03 | DD01 | |
| DB02 | AS03 | DD01 | |
| DB02 | DA02 | DD02 | |
| DB02 | DA03 | DD02 | |
| DB02 | AS03 | DA03 | |
| DB02 | DA02 | DD01 | |
| DB02 | DD01 | DD02 | |
| DB02 | DA03 | DA02 | |
| DB02 | AS03 | DA02 | |
| VA03 | VD03 | PVNL | Group 2 |
| VB02 | VD03 | PVNL | |
| VB02 | VA03 | PVNL | |
| VA03 | VD03 | VB02 | |
| SMDDL | SMDDR | SMBDL | Group 3 |
| DA06 | AVBR | AVBL | Group 4 |
| DA06 | AVBR | VA11 | |
| DA06 | DVB | AVBR | |
| VB08 | DA06 | AVBR | |
| VB10 | DA06 | AVBR | |
| VD09 | DA06 | AVBR | |
| VD10 | DA06 | AVBR | |
| VD11 | DA06 | AVBR | |
| VA09 | DA06 | AVBR | |
| VB09 | DA06 | AVBR | |
| DB04 | AVBR | AVBL | |
| DB04 | AVBR | VA11 | |
| DB04 | DVB | AVBR | |
| DB04 | VB08 | AVBR | |
| DB04 | VB10 | AVBR | |
| DB04 | VD09 | AVBR | |
| DB04 | VD10 | AVBR | |
| DB04 | VD11 | AVBR | |
| VA09 | DB04 | AVBR | |
| VB09 | DB04 | AVBR | |
| PDEL | VB10 | AVFR | |

**Table S5**: **Neuronal components of the triple ablation predictions.** Synthetic essential triplets occur in four distinct groups (see Figure 4(a)).

| Threshold $\tau$ (# synapses) | | | |
|---|---|---|---|
| 1 | 2 | 3 | 4 |
| *AS08* | *AS08* | *AS08* | *AVAL* |
| *AS09* | *AS09* | *AS09* | *AVAR* |
| *AS10* | *AS10* | *AS10* | *DA07* |
| *AS11* | *AS11* | *AS11* | *DA08* |
| *DA07* | *DA07* | *AVAL* | *DA09* |
| *DA08* | *DA08* | *AVAR* | *DB05* |
| *DA09* | *DA09* | *DA07* | *DD04* |
| *DB05* | *DB05* | *DA08* | *DD05* |
| *DB06* | *DB06* | *DA09* | *DD06* |
| *DB07* | *DB07* | *DB05* | *PDA* |
| *DD04* | *DD04* | *DB06* | *PVPL* |
| *DD05* | *DD05* | *DB07* | *VA11* |
| *DD06* | *DD06* | *DD04* | *VA12* |
| *PDA* | *PDA* | *DD05* | *VB07* |
| *PDB* | *PDB* | *DD06* | *VB08* |
| | | *PDA* | *VB09* |
| | | *PDB* | *VB10* |
| | | *VA09* | *VB11* |
|  |  | VA10 | VD08 |
|  |  | VA11 | VD09 |
|  |  | *VA12* | VD10 |
|  |  | VB07 | VD11 |
|  |  | VB08 | *VD12* |
|  |  | VB09 | *VD13* |
|  |  | VB10 |  |
|  |  | *VB11* |  |
|  |  | VD08 |  |
|  |  | VD09 |  |
|  |  | VD10 |  |
|  |  | VD11 |  |
|  |  | *VD12* |  |
|  |  | *VD13* |  |

**Table S6**: **Neuronal components of the single ablation predictions for a recently reanalysed wiring diagram.** Those in *italics* match the findings in the original connectome as per Table S3. All ablations are predicted to lead to the loss of independent control over one muscle cell.